# Anchoring Magnetic Field in Turbulent Molecular Clouds

Hua-bai Li[1*], C. Darren Dowell[2], Alyssa Goodman[1], Roger Hildebrand[3] and Giles Novak[4]

[1]*Harvard-Smithsonian Center for Astrophysics, 60 Garden Street, MS-78, Cambridge, MA 02138*

[2]*Division of Physics, Mathematics, and Astronomy, California Institute of Technology, MS 320-47, 1200 East California Boulevard, Pasadena, CA 91125; also Jet Propulsion Laboratory*

[3]*Enrico Fermi Institute and Department of Astronomy and Astrophysics, University of Chicago, 5640 South Ellis Avenue, Chicago, IL 60637; also Department of Physics*

[4]*Department of Physics and Astronomy, Northwestern University, 2145 Sheridan Road, Evanston, IL 60208*

## Abstract

One of the key problems in star formation research is to determine the role of magnetic fields. Starting from the atomic inter-cloud medium (ICM) which has density $n_H \sim 1$ cm$^{-3}$, gas must accumulate from a volume several hundred pc across in order to form a typical molecular cloud. Star formation usually occurs in cloud cores, which have linear sizes below 1 pc and densities $n_{H_2} > 10^5$ cm$^{-3}$. With current technologies, it is hard to probe magnetic fields at scales lying between the *accumulation length* and the size of *cloud cores*, a range corresponds to many levels of turbulent eddy cascade, and many orders of magnitude of density amplification. For field directions detected from the two extremes, however, we show here that a significant correlation is found. Comparing this result with molecular cloud simulations, only the sub-Alfvénic cases result in field orientations consistent with our observations.

---

[*] hli@cfa.harvard.edu

# 1. Introduction

In a medium with sufficient ionization, such as the bulk of a molecular cloud, magnetic field and mass should be well coupled (flux freezing). A weak magnetic field (corresponding to super-Alfvénic turbulence) will be tangled by turbulent eddies and we should expect no correlation between the field orientations inside molecular clouds and those in the surrounding ICM. On the other hand, a strong magnetic field (sub-Alfvénic turbulence) can channel turbulent flows and preserve field orientation over a large range of scales. These effects have been studied via numerical simulations of magnetohydrodynamic turbulence (e.g., Ostriker, Stone, & Gammie 2001; Price & Bate 2008; Falceta-Goncalves, Lazarian & Kowal 2008).

To distinguish a molecular cloud between globally super- and sub-Alfvénic, one can map the field morphology of the bulk volume of the cloud, and compare this with simulations (Li et al. 2006; Novak, Dotson, & Li 2009). But such observations are time consuming even with the efficiency of state-of-art instruments, and it has not yet been possible to study a large sample of clouds in this way. Therefore our knowledge of cloud magnetic fields has been mostly concentrated on cloud cores, using field morphologies (e.g. Schleuning 1998; Girart, Rao, & Marrone 2006), Chandrasekhar-Fermi method (e.g. Crutcher et al. 2004) and Zeeman measurements (e.g. Troland & Crutcher 2008). Knowing field strength from cloud cores, however, tells us nothing about the global property of a cloud, because both super- and sub-Alfvénic cores can possibly develop from either super- or sub-Alfvénic clouds (e.g. Burkhart et al. 2009). Nevertheless, global field strength and geometry of a cloud are crucial initial conditions which can significantly influence the efficiency and rate of star formation (e.g. Price & Bate 2008).

Though we are blind to the magnetic fields from the majority of a cloud, we are capable to detect field orientations from the two ends of its density spectrum. In this report, we compare magnetic field directions in high-density, small-scale cores (pc to sub-pc scale) with those in the low-density, large-scale ICM (several hundred pc scale). If the field orientation is preserved over such a large range of scales, then the field tangling due to turbulence eddies cannot be severe. If the field orientation is correlated between these two very different density regimes, then the field must be dynamically significant in comparison to the mass accumulation agents, i.e. gravity and turbulence. The methods we use to determine field orientations are described in § 2. We present our result and compare it with molecular cloud simulations in § 3. In § 4, we address the effects of several uncertainties, followed by a summary in § 5.

## 2. Method

It is empirically found that the polarization direction, for both optical starlight and mm/sub-mm thermal emission, is correlated with magnetic field orientation (e.g. Berkhuijsen et al. 1964). Most theories of grain alignment (Lazarian 2000) suggest that the shortest axes of spinning (non-spherical) dust grains will be preferentially parallel to the magnetic field. As a result, dust thermal emission and starlight extinguished by dust will be polarized, respectively, perpendicular and parallel to the field. Due to sensitivity constraints, observations of polarized thermal mm/sub-mm emission have until now mainly probed the high-density regions of a molecular cloud ($A_V > 100$), which typically have spatial extent below one parsec. Optical polarimetry, on the other hand, is primarily sensitive to the field in the ICM ($A_V < 3$, e.g. Arce et al. 1998, Poidevin & Bastien 2006). The accumulation length for a typical molecular cloud in the Milky Way is several hundred parsecs (Williams, Blitz, & McKee 2000) and field directions on this scale can be studied using optical polarimetry (Heiles 2000; Zweibel & Heiles 1997). The density contrast between the sub-pc scales probed by thermal emission and the several-hundred-pc scales probed optically is typically more than ten thousand.

### 2.1 *Sub-mm Polarimetry Data*

The sub-mm polarimetry data we use here were obtained from two recently released archives (Dotson et al. 2009; Curran & Chrysostomou 2007) which respectively present data collected using the 350 micron Hertz polarimeter (Dowell et al.1998) on the Caltech Submillimeter Observatory (CSO), and the 850 micron SCU-POL polarimeter (Greaves et al. 2003) on the James Clerk Maxwell Telescope (JCMT). From these archives, we have selected those cores having at least five 3-σ detections of polarization. For cores observed by both Hertz and SCU-POL, we find good agreement at most positions. Thus, for simplicity, we use only the Hertz data for such cores.

### 2.2 *Optical Polarimetry Data*

For each of the selected sub-mm cores, we chose a spherical "ICM region" centered at the core, and we studied the ICM magnetic field in this region using data from the optical polarimetry catalog of Heiles (2000). Since the accumulation length for a GMC is about 400 pc (Williams, Blitz, & McKee 2000), and since many of our targets are dark cloud cores rather than GMC cores, our ICM regions should not be larger than 400 pc. On the other hand, to get enough stars for reliable statistics, the ICM regions should not be too small either. With these constraints in mind, we started by choosing ICM regions of 200 pc diameter. For cases where such regions contain fewer than five stars from the catalog, we expanded the diameter of the spherical ICM region to 400 pc. If this expanded region still contained fewer than five stars, then the corresponding core was eliminated from our study. Twenty-five cores made our final list (Table I), all within 2 kpc of the Sun.

### 2.3 *Means and Interquartile Ranges*

Also listed in Table I are the means and interquartile ranges (IQRs) of the field directions inferred from the core and ICM polarimetry data. In evaluating the mean field directions, the smallest angular difference calculable between any two "vectors" is used (which is equivalent to the "equal weight Stokes mean" used by Li et al. 2006); for example, we use 20° and -10° instead of 20° and 170°. This is appropriate because polarization "vectors" are actually headless, in that they are ambiguous by 180°.

The IQR is defined as the difference between the 25th and 75th percentiles of a distribution. When there are extreme values in a distribution or when the distribution is skewed, as are the field distributions of some of the ICMs in our sample, the IQR is superior to the standard deviation as a measure the of the spread. For a normal distribution, the IQR is about 1.34 times greater than the standard deviation.

### 3. Results

#### 3.1 *Correlation between Sub-mm and Optical Data*

Figure 1 shows sub-mm data from several cores in the Orion molecular cloud (OMC) region, together with the corresponding ICM results. The region is approximately 100 parsecs across (for an assumed distance of 450 pc). Within it, eight cores were mapped by Hertz at 350 μm, each with linear size of about 0.3 pc. It can be seen that even though the core separations exceed the core sizes by as much as a factor of 100, they are for the most part "*magnetically connected*", i.e. the cores' mean field directions are similar. Moreover, these directions are close to the mean field direction seen in the ICM for the OMC region (see Figure 1 and Table I).

The cores in the OMC region are not special cases. The mean field directions of the 25 cores in our sample ($B_{core}$) are plotted against those of the corresponding ICM regions ($B_{ICM}$) in Figure 2. The average IQR for the ICM regions ($IQR_{ICM}$) is about 52°, and this range is indicated in the figure using orange shading. Note that even if the ICM field directions were perfectly preserved during the processes of cloud and core formation, we should not expect | $B_{core}$ - $B_{ICM}$ | to be zero. This is because, given the very different scales of the cores and their corresponding ICM regions, one $B_{core}$ can only sample a small portion of the magnetic field in the ICM regions. So only half of the $B_{core}$ are expected to fall within 26° from the corresponding $B_{ICM}$ because of the $IQR_{ICM}$. Figure 2 shows that almost 70% of the $B_{core}$ are within 26° from $B_{ICM}$.

Another indication of a significant correlation is that nearly 90% of the $B_{core}/B_{ICM}$ pairs are more nearly parallel than perpendicular (i.e. nearly 90% have | $B_{core}$ - $B_{ICM}$ | < 45°). The probability for obtaining this correlation from two random distributions is less than $7\times10^{-5}$. It might be argued that this calculation is flawed since it ignores the fact that some of the cores in our sample belong to the same cloud, so their field directions might be correlated. However, note that this objection assumes that the field is sufficiently dynamically important to prevent field tangling by turbulence during the core formation process, which is the hypothesis we are currently testing.

It has been shown (e.g. Hang & Zhang 2007; Reid & Silverstein 1990) that the sign of the line-of-sight component of the Galactic magnetic field is preserved in molecular clouds, using the correlation between the Zeeman splitting data of masers and rotation measures of pulsars. Our polarimetry data not only show that the components in the plane of the sky are preserved, but also put constraints on star and cloud formation theories, as described in the following section.

### 3.2 Comparing with Simulations

Ostriker, Stone, & Gammie (2001) simulated molecular clouds with various Alfvénic Mach numbers ($M_A \equiv v_t/v_A$, where $v_t$ and $v_A$ are respectively turbulent and Alfvenic velocities), and the results (e.g. field orientations) were projected along lines of sight at various angles ($\theta$) from the mean field direction. They divided the simulations into three categories: strong, moderate, and weak magnetic fields, which are corresponding to $M_A \approx$ 0.7, 2, and 7 for cases shown in their Figure 24. In that figure, distributions of cloud field orientations are plotted with respect to the most-frequent directions. The observed data ($B_{core}$ and $B_{ICM}$) are from cloud cores and ICM, while the simulated field orientations are mostly from the "bulk cloud volume", which are the regions between cores and ICM. To compare the observations with the simulations, we need to make assumptions, which are based on the fact that $B_{core}$ and $B_{ICM}$ are highly correlated:

1. $B_{ICM}$ is the mean field orientation in the initial condition of cloud formation, and is close to the most-probable field direction in a cloud.
2. $B_{core}$ is a good representative of field directions in the bulk volume of a cloud.

With these assumptions, the offset of $B_{core}$ from $B_{ICM}$ is comparable to cloud field orientations with respect to the most-frequent directions. In Figure 3 are a histogram of $B_{core}$ - $B_{ICM}$ and two field distributions ($M_A \approx 0.7$ and 2; both with $\theta = 45°$) from Figure 24 of Ostriker et al. (2001).

The strong field case fits well to the central 80% (-40° − 40°) of the $B_{core}$ - $B_{ICM}$ histogram. The more significant "tails" of the observational data could be because some clouds have $M_A > 0.7$ and/or $\theta < 45°$. Some of the disagreement is no doubt due to factors that are not included in the simulations and due to the oversimplified assumptions. For example, the observed $B_{ICM}$ has a significant IQR, while the B field in the initial condition of the simulations is uniform. Also, the observed $B_{core}$ values may be significantly altered by stellar feedback (e.g. HII regions and outflows, which are not simulated) and possibly not correlated with the fields in the bulk cloud volumes anymore.

On the other hand, even though some random factors are not included in these simulations, the moderate field case ($M_A \approx 2$) can not produce a distribution as peaked as the $B_{core}$ - $B_{ICM}$ histogram, no matter how $\theta$ is adjusted (see Figure 24 of Ostriker et al. 2001 for other $\theta$'s). The kurtoses of $B_{core}$ - $B_{ICM}$ and the strong and moderate field distributions shown in Figure 3 are 3.7, 3.2, and 2.8, respectively. Falceta-Goncalves, Lazarian & Kowal (2008) presented similar simulations, and their super-Alfvénic ($M_A \approx 2$) case shows a random angle distribution (Figure 3).

### 4. Discussion

In this section, we discuss some uncertainties and why we believe that sub-mm polarimetry probes most of cloud fields along sight lines and that optical polarimetry probes the surrounding ICM.

### 4.1 *Probing into Deep Clouds*

The lowest polarization fraction from a cloud core usually happens at or close to the total flux peak (see Figure 4 for an example from DR 21). This phenomenon is called a "polarization hole". Low grain alignment efficiencies in high-density environments have been proposed to explain polarization holes. Some (e.g. Padoan et al. 2001) suggested that grains can not be aligned for $A_v > 3$ mag. If this were true, sub-mm polarimetry definitely could not probe the cloud cores in our sample, whose typical $A_v$ is above $10^3$ mag[*]. However, in the following we argue that a low grain alignment efficiency is not the only possible reason for polarization holes, and that a cut-off at $A_v = 3$ mag is not supported by observations.

The Chandrasekhar-Fermi method is widely used to estimate magnetic field strength ($B$) in molecular clouds using cloud density ($\rho$), turbulent velocity dispersion ($\sigma$), and field deviation ($\alpha$) from a mean direction:

$$\frac{B(G)}{\sqrt{\frac{4\pi}{3}\rho(g/cm^3)}} = \frac{\sigma(cm/\sec)}{\alpha(rad)} \quad (1)$$

Here we use this relation slightly differently. Given a fixed ratio of $B$ to $\sqrt{\rho}$, $\alpha$ is proportional to $\sigma$, which roughly increases with column density (e.g. Pillai et al. 2006; Burkhart et al. 2009)[♦]. Thus $\alpha$ can also increase with flux. An increase of $\alpha$ within a telescope beam will decrease the polarization fraction as shown in the following. Assume that the degree of polarization is a constant $P_0$ for $\alpha = 0$ (i.e. uniform grain alignment efficiency and field orientation). For simplicity, we also assume that field orientations in a

---

[*] NGC 1333 has the 25th percentile of the 350 μm peak fluxes of our sample cores. The calculation near the end of this section indicates that the peak $A_v$ of NGC 1333 is around 1500 mag (based on the JCMT data).

[♦] For starless cores, $\sigma$ can decrease as flux increases (Goodman et al. 1998; Pillai et al. 2006). All the cores in our sample have embedded stars; polarimetry of starless cores is very difficult with the sensitivities of current polarimeters.

telescope beam are evenly distributed from $\theta - \alpha$ to $\theta + \alpha$, with a mean field direction $\theta$ and angle deviation $0 < \alpha < \pi/2$. The mean Stokes parameters of this group of vectors are:

$$q = \frac{P_0}{2\alpha} \int_{-\alpha}^{\alpha} \cos 2(\theta + \phi) d\phi, \text{ and } u = \frac{P_0}{2\alpha} \int_{-\alpha}^{\alpha} \sin 2(\theta + \phi) d\phi, \quad (2)$$

and the mean polarization fraction can be shown as

$$\overline{P} = \sqrt{q^2 + u^2} = P_0 \frac{\sin 2\alpha}{2\alpha} = P_0 \frac{\sin 2\sigma/c}{2\sigma/c}, \quad (3)$$

where $c$ is $B/\sqrt{4\pi\rho/3}$. Zeeman measurements (e.g. Vallée 1997, Crutcher 1999) show that $c$ can be fitted to within the errors by a constant. Though this correlation is approximate because only one component of $B$ is measured (and possibly some other reasons; see Basu 2000 and Myers & Goodman 1988), it should still be good enough to reveal a trend of $\overline{P}$ versus $\sigma$, i.e. a polarization hole. Overlapped on the $\overline{P}$ − flux plot of DR 21 in Figure 4 is a $\overline{P} - \sigma$ plot from the same sky positions. We note that the $\sigma$ data in Figure 4 is based on HCN (4-3) lines (Kirby 2009), which could be optically thicker than the 350 μm dust emission, from which the polarimetry data is derived. So the correlation may change when optically thinner lines are used for $\sigma$, and the trend may be enhanced. Kirby (2009) estimated the sky-component of the mean field in this region as 2.5 mG, and Jakob et al. (2007) estimated the n(H$_2$) as $10^6$/cm$^3$. Based on these two parameters, equation (3) is also plotted in Figure 4, assuming a mean molecular mass of 2.3 and $P_0$ = 3 %, which is estimated by the mean of the polarizations from the lowest 10 % of the flux with 3-sigma polarization detections from DR 21 Main (Dotson 2009).

An interesting fact is that sub-mm polarization holes show at very different scales: ~10 pc (Li et al. 2006), 0.1-1 pc (this work), and below 0.1 pc (e.g. Girart et al. 2006). The average $A_v$'s from these scales are different in order of magnitude, so switching off grain alignment above one particular $A_v$ can not explain polarization holes from all these scales.

Even when one ignores the effect of velocity dispersion discussed above and charges polarization holes completely to grain alignment efficiency, $A_v$ = 3 mag is still too low for a cut-off on alignment efficiency to explain many observations. In our sample, the mean polarization is 2.55 % for the 955 3-sigma detections from the Hertz archive. With the maximum possible polarization around 10 % (Hildebrand & Dragovan 1995) and the typical $A_v$ for our samples over $10^3$ mag, 2.55 % requires grains to be aligned to as deep as $A_v$ = 255/2 mag, assuming $\alpha$ = 0 and that the polarized mass is evenly distributed to the near and far sides of a core. So, even $A_v$ = 125 mag is a conservative estimate of the cut-off point. Interferometer polarimetry also needs to be explained. Take NGC 1333 IRAS 4A for an example, the 800 μm flux from the central 33 arcsec$^2$ region is about 6 Jy according to the SMA observation (Girart et al. 2006) and is approximately 11 Jy from the JCMT (Sandell et al. 1991; scaled down from 12.9 Jy over 39 arcsec$^2$). The 5 Jy missing flux filtered out by the SMA is corresponding to a fore/background column density $N(H_2)$ around $7\times10^{23}$ cm$^{-2}$ (Girart et al. 2006), which is about $A_v$ = 700 mag (Harjunpää, Lehtinen & Haikala 2004). Therefore the grains are aligned deeper than $A_v$ = 350 mag.

Our goal is to distinguish clouds between globally super- and sub-Alfvénic by comparing the field orientations from the two ends of the density spectrum within a bulk cloud volume (§ 3.2). For this purpose, knowing that grains are aligned to at least hundreds of magnitude in $A_v$ is good enough, because even for a giant molecular cloud, the typical $A_v$ of the bulk volume is in the order of 10 mag (McKee & Ostriker 2007).

### 4.2 *Optical Foreground*

Any single polarization measurement gives the mean field direction along the line of sight, weighted by the density and the alignment efficiency of dust particles. In principle, the optical polarimetry data will only reliably give the field orientation in each ICM region if we first correct these data for the foreground contribution to the polarization.

In this work we assume that the foreground effects are not dominant, and make no correction for foreground polarization. Our reasons for ignoring foreground effects are twofold: First, accurate foreground subtraction requires one to identify a significant number of "foreground stars" lying very close in the sky to each "background star" whose polarization one is attempting to correct. Starting from the limited number of stars in the Heiles catalog, this is often not possible. For example, in our earlier work (Li et al. 2006) we searched for foreground stars for each background star, and we imposed the following restrictions on foreground correction: (1) there should be more than three foreground stars within 1.5° of each background star, and (2) they should lie approximately one accumulation length (400 pc) closer to us in comparison with the background star. With these restrictions, only ten out of 77 nearby regions studied were found to have five or more background stars that could be corrected for foreground polarization.

Our second reason for ignoring foreground subtraction is that during the course of carrying out the work reported in Li et al. (2006), we noticed that foreground subtraction typically tends to reduce the polarization fraction while not introducing significant changes to the polarization direction. This effect agrees with Figure 1 of Andreasyan & Makarov (1989), where the degree of optical polarization for stars located within 2 kpc tends to increase with distance even though polarization direction stays nearly constant, for most sight-lines.

Most importantly, our conclusion that core fields are correlated with ICM fields does not depend on the assumption that foreground polarization is negligible. To see this, assume that indeed our $B_{ICM}$ values are primarily measuring the foreground polarization. In this case, Figure 2 implies that the $B_{core}$ values are correlated with ICM fields over distances larger than the accumulation length. It is hard to imagine how this could be possible unless the $B_{core}$ values were also correlated with the ICM fields within an accumulation length surrounding the cloud.

In other words, foreground effects merely add noise to Figure 2. If one assumes that the foreground effects must be important, then this noise must be large. In this case the underlying correlation between the $B_{core}$ values and the ICM fields within an accumulation length surrounding the cloud must be even stronger, so that one can obtain the observed correlation of Figure 2 despite the large noise level.

## 5. Summary

We compare the field directions inferred from the sub-mm polarimetry data from 25 cloud cores (diameter < 1 pc, $A_v$ > 1000 mag) with field directions inferred from optical polarimetry data from their surrounding inter-cloud media (diameter > 200 pc, $A_v$ < 3 mag) and see a significant correlation (Figure 2). Comparing with simulations, a globally super-Alfvénic cloud does not fit into this picture (Figure 3).

We appreciate the referee, whose comments have made our article much better. We are grateful to Sridharan K. Tirupati, Zhi-Yun Li, Paolo Padoan, Telemachos Ch. Mouschovias, Yasuo Fukui, Alex Lazarian and John Scalo for insightful discussions and comments. H. Li appreciates the helps from Tingting Wu on collecting and preliminarily analyzing the optical data. We thank Scott Paine and Robert Kimberk for proofreading many versions of the manuscript. H. Li's research is funded through a postdoctoral fellowship from the Smithsonian Astrophysics Observatory. R. Hildebrand acknowledges NSF grant AST-0505124 for support of the submillimeter observations. The Caltech Submillimeter Observatory is funded through the NSF grant AST-0540882.

**References**

Andreasyan, R. R. & Makarov, A. N. 2000, Ap 31, 560

Basu, S., 2000, ApJ, 540, L103

Berkhuijsen, E. M., Brouw, W. N., Muller, C. A. & Tinbergen, J. 1964, Bul. Astro. Net. 17, 465

Burkhart, B., Falceta-Gonçalves, D., Kowal, G., & Lazarian, A., 2008, ApJ, 693, 250

Curran, R. L. & Chrysostomou, A. 2007, MNRAS 382, 699

Crutcher, R., 1999, ApJ, 520, 706

Crutcher, R., Nutter, D., Ward-Thompson, D., & Kirk, J., 2004, ApJ, 600, 279

Dotson, J. L., Davidson, J. A., Dowell, C. D., Kirby, L., Hildebrand, R. H. & Vaillancourt, J. E. 2009, submitted to ApJS

Dowell, C. Darren; Hildebrand, Roger H.; Schleuning, David A.; Vaillancourt, John E.; Dotson, Jessie L.; Novak, Giles; Renbarger, Tom; Houde, Martin, 1998, ApJ, 504, 588

Falceta-Goncalves, D., Lazarian, A. & Kowal, G. 2008, ApJ, 679, 537

Arce, H., Goodman, A., Bastien, P., Manset, N., & Sumner, M., 1998, ApJ, 499, 93

Girart, J., Rao, R., & Marrone, D., 2006, Sci, 313, 812

Goodman, A, Barranco, J, Wilner, D., & Heyer, M., 1998, ApJ, 504, 223

Greaves, J. S.; Holland, W. S.; Jenness, T.; Chrysostomou, A.; Berry, D. S.; Murray, A. G.; Tamura, M.; Robson, E. I.; Ade, P. A. R.; Nartallo, R.; Stevens, J. A.; Momose, M.; Morino, J.-I.; Moriarty-Schieven, G.; Gannaway, F., & Haynes, C. V. 2003, MNRAS, 340, 353

Han, J. & Zhang, J. 2007, A&A, 464, 609

Harjunpää, P., Lehtinen, K., & Haikala, L. K., 2004, A&A, 421, 1087

Heiles, C. 2000, AJ, 119, 923

Hildebrand, R. & Dragovan, M., 1995, ApJ, 450, 663

Jakob, H, Kramer, C, Simon, R., Schneider, N., Ossenkopf, V., Bontemps, S., Graf, U. & Stutzki, J., 2007, A&A, 461, 999.

Kirby, L., 2009, ApJ, 694, 1056

Lazarian, A. 2000, ASPC, 215, 69

Li, H.; Griffin, G. S.; Krejny, M.; Novak, G.; Loewenstein, R. F.; Newcomb, M. G.; Calisse, P. G. & Chuss, D. T. 2006, ApJ, 648, 340.

McKee, C. & Ostriker, E., 2007, ARA&A, 45, 565

Myers, P. & Goodman, A., 1988, ApJ, 329, 392

Neugebauer, G.; Habing, H. J.; van Duinen, R.; Aumann, H. H.; Baud, B.; Beichman, C. A.; Beintema, D. A.; Boggess, N.; Clegg, P. E.; de Jong, T.; Emerson, J. P.; Gautier, T. N.; Gillett, F. C.; Harris, S.; Hauser, M. G.; Houck, J. R.; Jennings, R. E.; Low, F. J.; Marsden, P. L.; Miley, G.; Olnon, F. M.; Pottasch, S. R.; Raimond, E.; Rowan-Robinson, M.; Soifer, B. T.; Walker, R. G.; Wesselius, P. R.; & Young, E. 1984, ApJ, 278, L1

Novak, G., Dotson, J. L., & Li, H. 2009, ApJ, 695, 1362

Ostriker, E. C., Stone, J. M., & Gammie, C. F. 2001, ApJ, 546, 980

Padoan, P., Goodman, A. Draine, B., Juvela, M., Nordlund, & Å. Rögnvaldsson, Ö. 2001, ApJ, 559, 1005

Poidevin, F., & Bastien, P., 2006, ApJ, 650, 945

Reid, M. & Silverstein, E., 1990, ApJ, 361, 483

Price, D. & Bate, M. 2008, MNRAS, 385, 1820

Schleuning, D., 1998, ApJ, 493, 811

Pillai, T., Wyrowski, F., Carey, S. J., & Menten, K. M., 2006, A&A, 450, 569

Troland, T. & Crutcher, R., 2008, ApJ, 680, 457

Vallée, J., 1997, Fund.Cosmic Phys., 19, 1

Williams, J. P., Blitz, L., & McKee, C. F. 2000, Protostars and Planets IV, Mannings, A. P. Boss, & S. S. Russell (Tucson: Univ. Arizona Press), 97

Zweibel, E. & Heiles, C. 1997, Nature, 385, 131

**Table I**  A summary of the 25 selected molecular clouds

| Cloud Name | Position | | | Sub-mm inferred B | | | Optical inferred B | | |
|---|---|---|---|---|---|---|---|---|---|
| | $l$ (°)[a] | $b$ (°)[a] | $d$(pc) | Mean(°)[b] | IQR(°) | # det. | Mean(°)[b] | IQR(°) | # det. |
| GGD 27[c] | 10.8 | -2.6 | 1700 | 3.3 | 60.6 | 23 | 37.3 | 66.3 | 6 |
| M17 | 15.0 | -0.7 | 1500 | 77.4 | 45.4 | 129 | 65.2 | 79.0 | 13 |
| IRAS 20126+4104[c] | 78.0 | 4.0 | 1700 | 55.7 | 28.4 | 18 | 26.2 | 66.0 | 22 |
| GL2 591[c] | 78.9 | 0.7 | 1500 | 93.2 | 42.4 | 37 | 24.0 | 76.0 | 15 |
| S140[c] | 106.8 | 5.3 | 900 | 46.9 | 63.7 | 96 | 57.9 | 43.0 | 17 |
| Cep A[c] | 109.9 | 2.1 | 730 | 33.7 | 26.4 | 106 | 49.9 | 35.0 | 14 |
| NGC 1333 | 158.4 | -20.6 | 350 | 176.2 | 16.1 | 8 | 131.5 | 54.3 | 13 |
| Mon OB1 IRAS 12 | 203.2 | 2.1 | 800 | 51.6 | 15.8 | 26 | 178.2 | 32.4 | 5 |
| NGC 2264 | 203.3 | 2.1 | 800 | 167.1 | 14.9 | 18 | 178.2 | 32.4 | 5 |
| NGC 2071 | 205.1 | -14.1 | 390 | 65.0 | 35.0 | 6 | 69.9 | 44.7 | 95 |
| NGC 2068 LBS10 | 205.4 | -14.3 | 400 | 42.0 | 36.6 | 20 | 69.7 | 45.5 | 91 |
| NGC 2024 | 206.5 | -16.4 | 420 | 52.9 | 21.2 | 26 | 66.9 | 40.0 | 92 |
| OMC-3 | 208.0 | -17.0 | 450 | 48.5 | 15.2 | 44 | 70.6 | 51.8 | 93 |
| OMC-2 | 208.0 | -19.0 | 450 | 66.2 | 37.8 | 18 | 70.4 | 53.3 | 93 |
| IRAS 05327-0457 | 208.6 | -19.2 | 450 | 50.7 | 16.8 | 18 | 68.5 | 52.2 | 92 |
| OMC-1 | 209.0 | -19.4 | 450 | 114.9 | 36.5 | 351 | 68.5 | 52.2 | 92 |
| OMC-4 | 209.2 | -19.5 | 450 | 76.2 | 21.1 | 16 | 68.5 | 52.2 | 92 |
| Mon R2 | 213.7 | -12.6 | 830 | 126.3 | 42.0 | 51 | 108.6 | 56.0 | 10 |
| GGD 12 | 213.9 | -11.8 | 1000 | 178.8 | 27.3 | 17 | 179.3 | 66.4 | 20 |
| NGC 6334 V | 351.2 | 0.7 | 1700 | 0.6 | 31.7 | 8 | 155.2 | 39.0 | 10 |
| NGC 6334 A | 351.3 | 0.7 | 1700 | 157.6 | 15.9 | 49 | 155.9 | 40.3 | 9 |
| NGC 6334 I | 351.4 | 0.7 | 1700 | 127.3 | 22.0 | 33 | 147.6 | 35.2 | 8 |
| ρ Oph | 353.1 | 16.9 | 130 | 70.0 | 19.4 | 9 | 89.9 | 60.1 | 372 |
| IRAS 16293 | 353.9 | 15.8 | 120 | 176.6 | 27.9 | 7 | 86.4 | 59.4 | 389 |
| Rcr A[c] | 359.9 | -17.9 | 130 | 91.0 | 33.5 | 26 | 88.6 | 64.1 | 397 |

[a] The positions are shown in Galactic coordinates.

[b] The inferred B directions are measured from celestial north-south in J2000 coordinates, increasing counterclockwise. All the angles are shown in degrees.

[c] From Curran & Chrysostomou (2007). The rest are from Dotson et al (2009)

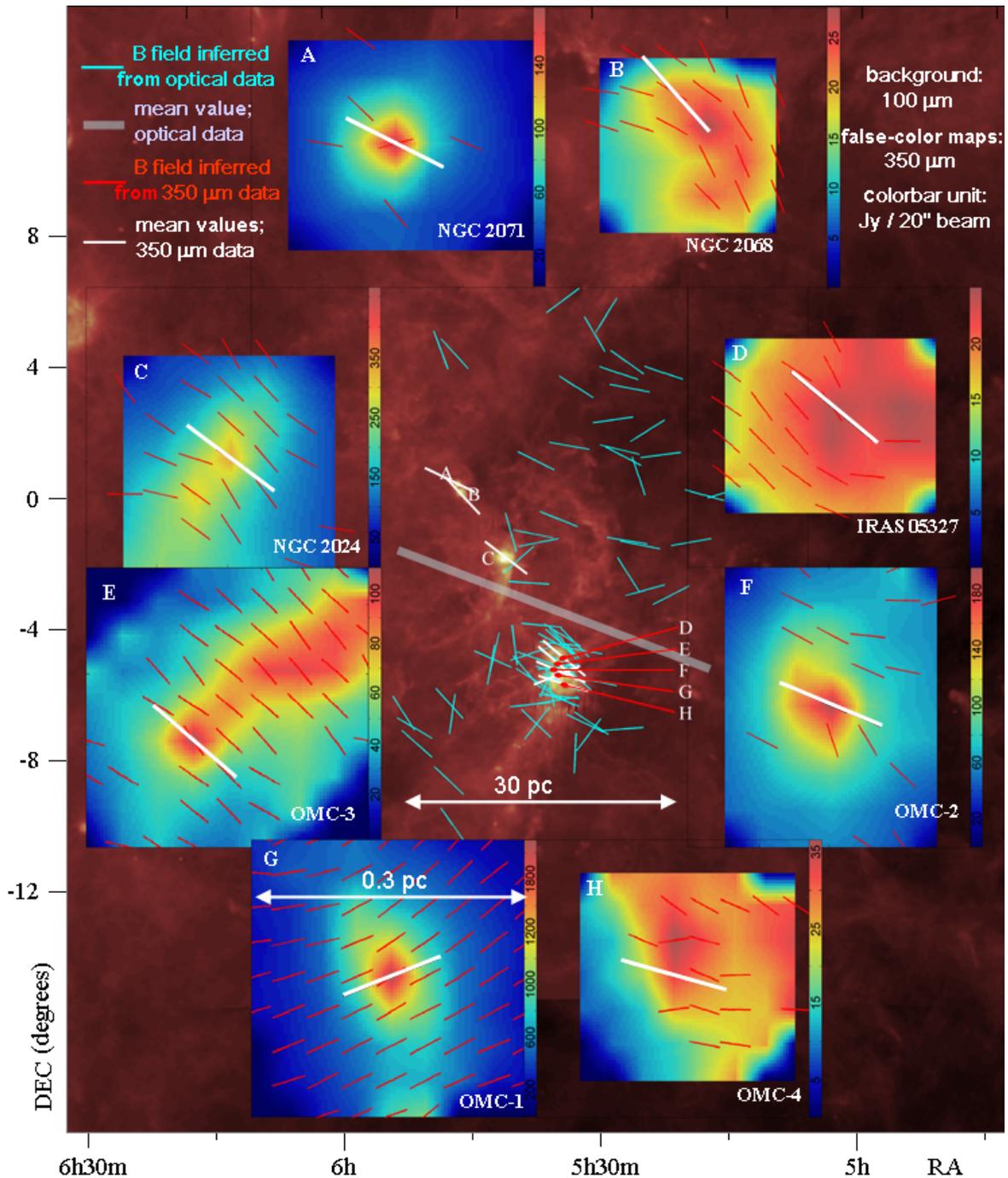

**Figure 1. Magnetic fields in the Orion molecular cloud region**. The background image shows the IRAS (Neugebauer et al. 1984) 100 μm map in logarithmic scale. We superpose on this map the magnetic field directions inferred from optical data (blue vectors), and the mean of all the optical data is shown as the thick gray vector. The Hertz polarimeter (Dotson et al. 2009) at the Caltech Submillimeter Observatory mapped eight clouds (see labels A through H on the IRAS map) in this region at 350 μm with 20" resolution, and these CSO results are shown as insets, using red vectors on individual false-color intensity maps. The mean direction of all the 350 μm polarization detections from a given core is shown as a white vector superposed on each core's map, and these white vectors are also plotted on the IRAS 100 μm map. All the false-color Hertz intensity maps are plotted to the same scale: 140 arcseconds across (approximately 0.3 pc). Note that the spatial scales and mass densities are very different between the regions probed by the two wavelengths, but the field orientations are very similar.

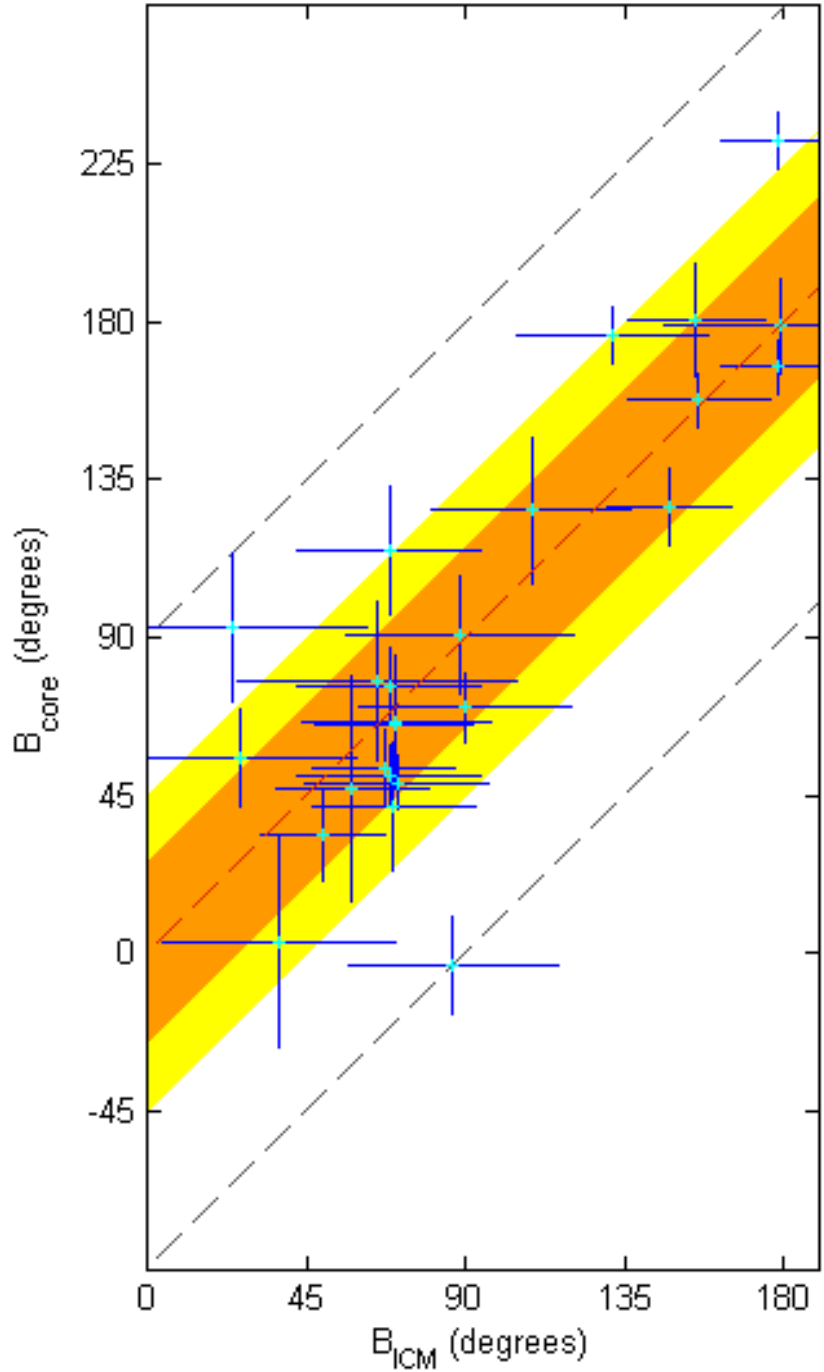

**Figure 2. Correlation between $B_{core}$ and $B_{ICM}$.** $B_{core}$ and $B_{ICM}$ are, respectively, the mean field direction in each cloud core and that in the corresponding ICM region surrounding that core. The directions are measured from north-south in J2000 coordinates, increasing counterclockwise. Each cloud core is within 2 kpc from the Sun and contains at least five submm polarization detections of 3-$\sigma$ significance. Each corresponding ICM region contains at least five optical stellar polarization detections of 3-$\sigma$ significance. The blue bars indicate the interquartile ranges (IQRs) of the polarization angle distributions. The mean of the IQRs from all the ICM regions is approximately 52° (see orange area). Note that about 70% of the core/ICM pairs deviate from perfect parallelism by less than 26°. For nearly 90% of the cores, the mean field is more nearly parallel than perpendicular to the mean field of the surrounding ICM (orange and yellow regions together).

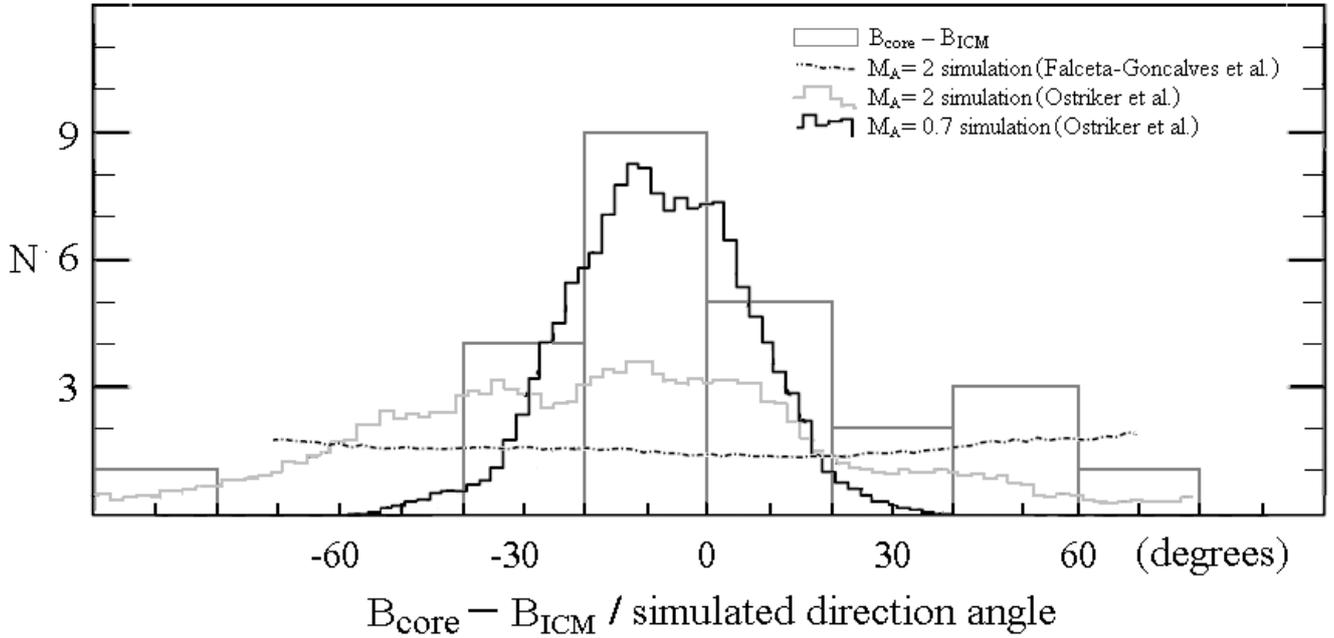

**Figure 3 Comparisons between the observations and the simulations**. The coarse histogram is from the observed offset of $B_{core}$ from $B_{ICM}$. The black and gray distributions (fine histograms) of field orientation are from simulations by Ostriker et al. (2001) and plotted with respect to the most-frequent direction. The black distribution is from a sub-Alfvénic case ($M_A \approx 0.7$), while the gray distribution is super-Alfvénic ($M_A \approx 2$); both results are projected along a line of sight 45° from the mean field direction. Another super-Alfvénic simulation ($M_A \approx 2$ with a line of sight 90° from the mean field direction) from Falceta-Goncalves et al. (2008) obtained a random distribution shown as the dashed line. The simulated distributions have been rescaled for visual comparisons with the shape of the $B_{core} - B_{ICM}$ histogram.

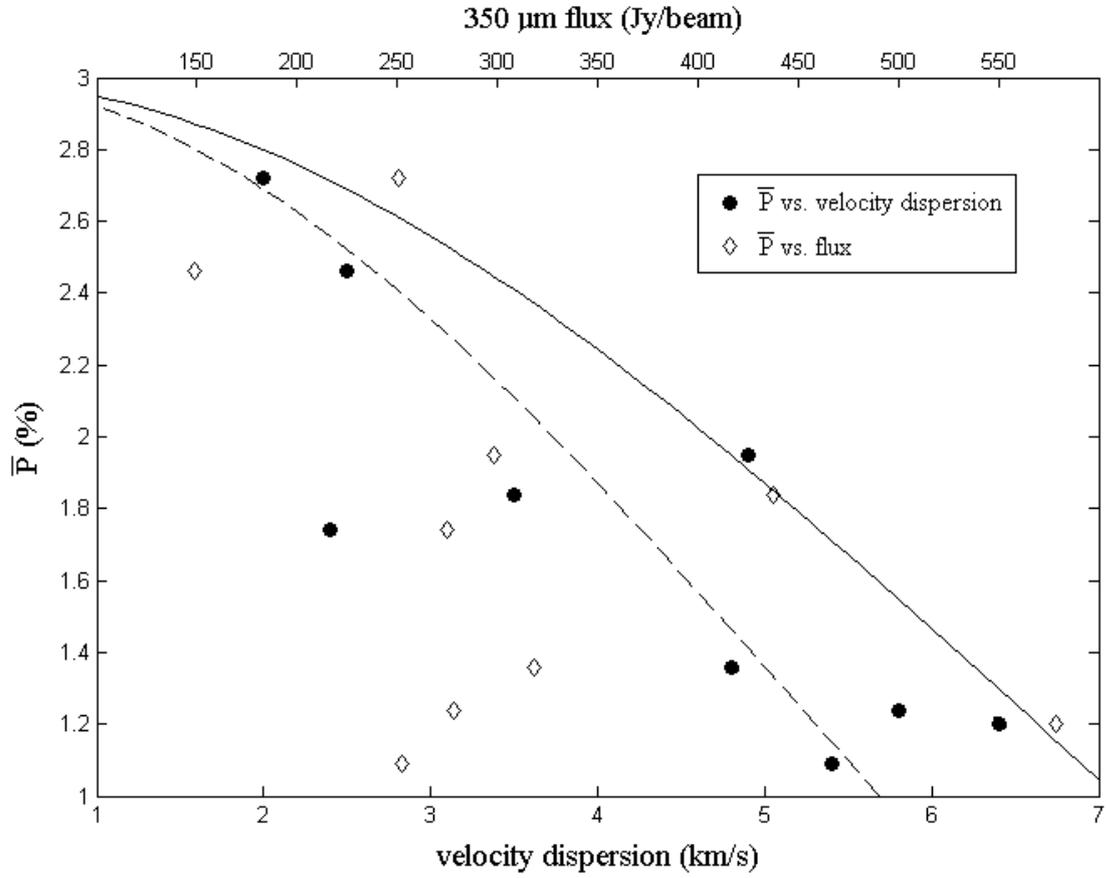

**Figure 4 The Polarization hole from DR 21 Main.** The 350 μm flux and HCN (4-3) velocity dispersion data ($\sigma$) are from Kirby (2009). The 350 μm polarization fractions ($\bar{P}$) are from the Hertz archive (Dotson et al. 2009). The solid line is from our model of the $\bar{P}$ - $\sigma$ relation (equation (3)), assuming B = 2.5 mG (Kirby 2009), n($H_2$) = $10^6$/cm$^3$ (Jakob et al. 2007), and $P_0$ = 3 %; see section 4.1 for detail. The B = 2 mG case is also plotted (dashed line).